\documentclass[prl,twocolumn,floatfix,showpacs,showkeys]{revtex4} 

\usepackage[first,draft]{draftcopy}
\usepackage{latexsym,amssymb,amsmath,amsfonts}
\usepackage{graphicx}
\usepackage{bm}
\newcommand{\beq}{\begin{equation}}
\newcommand{\eeq}{\end{equation}}
\newcommand{\bc}{\begin{center}}
\newcommand{\ec}{\end{center}}
\newcommand{\eeqa}{\end{eqnarray}}
\newcommand{\beqa}{\begin{eqnarray}}
\newcommand{\no}{\noindent}

\newcommand{\ra}{\rightarrow}

\newcommand{\ga}{\gamma}

\newcommand{\de}{\delta}

\newcommand{\si}{\sigma}

\newcommand{\om}{\omega}
\newcommand{\ed}{\end{document} }

\begin{document}

\title{Spin spin interaction and the relativistic Rabi formula}
\author{Richard T. Hammond}

\email{rhammond@email.unc.edu }
\affiliation{Department of Physics\\
University of North Carolina at Chapel Hill\\
Chapel Hill, North Carolina and\\
Army Research Office\\
Research Triangle Park, North Carolina}

\date{\today}


\begin{abstract}
The interaction of an electromagnetic wave with spin (polarized light) and an electron is computed. Specifically the spin flip probability is computed using the Dirac equation for an electron trapped in a uniform magnetic field.
\end{abstract}

\maketitle
   
\section{Introduction}

The spin state of an electron is becoming increasing important as spintronics continues to develop and quantum information seeks long lived cubit states. For these reasons it is important to know the effect of an electromagnetic wave on the spin state. In particular we consider an electron trapped in a uniform magnetic field. Now we assume an electromagnetic pulse is incident on the electron and ask, what is the spin flip probability.

The spin flip probability was recently computed for an electron in hydrogen,\cite{davis} but this was an interaction involving the angular momentum of the electron. Spin flips were also studied from the Kapitza-Dirac mechanism.\cite{ahrens} Here we would like to consider the interaction of a polarized electromagnetic wave and the intrinsic spin of the electron. Since polarization arises from the intrinsic spin of the electromagnetic field, this may be properly called a spin spin interaction. A brief review of the literature may be found in Ref. \cite{hammond}.

Following this thought, one might be tempted to ask, what is the spin flip probability of a free electron due to an incident electromagnetic wave? To pursue this, the direction of the spin of the electron must be known. In order to have a fiduciary field against we measure the spin, we  consider the interaction of an incident wave with an electron trapped in a magnetic field.

The exact nature of this external magnetic field depends on the particular problem involved, but we shall consider the case of a uniform magnetic field. This is approximately true in magnetic traps if the particle stays near the center.

Let us consider the Dirac spinor, which is the true description of an electron. As is seen below, there is an array of terms that may arise, and therefore, in order to assess all possible spin field interactions, it is essential to use the Dirac solutions. Natural units, with $\hbar = 1$ and $c=1$ are used until the end, where cgs is adopted.

The solution for an electron trapped in uniform magnetic fields was solved long ago,\cite{huff} here I will highlight the results
The wave function for an electron trapped in a uniform magnetic field in the $z$ direction is

\beq
\psi_n=C_ne^{-i(E_nt-p_x^ix-p_z^fz)}e^{-\xi^2/2}u_n
\eeq

\no where the momentum terms are eigenvalues, 

\beq
C_n^2=\frac{\sqrt{eB}(E_n+m)}{8L_xL_zE_n}
\eeq
and

\beq
\xi=\sqrt{eB}y -\frac{p_x}{\sqrt{eB}}
.\eeq
\no  We assume the final state is spin up and the initial state is spin down, so that,

\beq
u_f=\left(
\begin{array}{cccc}
h_{n-1} &\\
0 &\\
p_z^fh_{n-1}/(E_f+m) &\\
-\sqrt{2neB}h_n/(E_f+m) &\\
\end{array}\right)
\eeq
and

\beq
u_i=\left(
\begin{array}{cccc}
0 &\\
h_n &\\
-\sqrt{2neB}h_{n-1}/(E_i+m)  &\\
-p_z^ih_n/(E_i+m) &\\
\end{array}\right)
\eeq

\no where $h_n=N_nH_n$ where $H_n$ are the Hermite polynomials $N_n=1/\sqrt{2^nn!\sqrt{\pi}}$, and the energy eigenvalue $n$ is given by $n=0,1,2,...$ with the convention if the subscript of the Hermite polynomial is negative, then it is zero.

It is assumed that we have two dimensional box normalization, the sides being $L_x$ and $L_z$. The functions $u$ are functions of $y$ which goes from minus to plus infinity. It may be noted that the gauge freedom in the choice of the electromagnetic potential translate to a freedom in the choice of $y$ or $x$, or a combination of the two.

It is assumed the electron interacts with an incoming electromagnetic field described by the four potential $A_\mu$. Now we are in a position to consider 
 the transition amplitude, defined by

\beq\label{ta}
S_{fi}=-ie\int d^4 x\overline\psi_f\ga^\si A_\si\psi_i
\eeq

\no where $f$ and $i$ denote final and initial states and
$A_\si$ is the electromagnetic potential.

Normally the final and initial states are free particle wave functions, but that is not sensible here. We are considering the scattering of a bound state to another bound state, so the wave functions should be those of an electron in a magnetic  field, which is assumed to be constant.

To begin let us write (\ref{ta}) as

\beqa\label{ta2}
S_{fi}=
-\frac{iC_fC_i}{L}\int d^4x 
e^{i(E_ft-p_x^fx-p_z^fz)}e^{-\xi^2}\\ \nonumber
\times M_{fi}e^{-i(E_it-p_x^ix-p_z^iz)}
\eeqa
where
$L=\om/eE$ and
the matrix element is

\beq
M_{fi}=u_f^\dag\ga^0\ga^\si A_\si u_i
,\eeq
which becomes, assuming $n\ra n+1$,

\begin{widetext}
\beqa\label{rich}
M=A_0 \left(\frac{\sqrt{2}
   \sqrt{B (n+1) e} h_n h_{n+1}
   p_z^i}{(E_f+m)(E_i+m)}
   -\frac{\sqrt{2}
   \sqrt{B n e} h_{n-1} h_n
   p_z^f}{(E_f+m)
   (E_i+m)}\right)\\ \nonumber
+A_1
   \left(\frac{h_n^2
   p_z^f}{E_f+m}-\frac{h_n^2
   p_z^i}{E_i+m}\right)
+A_2 \left(\frac{i h_n^2
   p_z^i}{E_i+m}
   -\frac{i h_n^2
   p_z^f}{E_f+m}\right)\\ \nonumber
+A_3 \left(\frac{\sqrt{2} \sqrt{B
   (n+1) e} h_n
   h_{n+1}}{E_f+m}-\frac{\sqrt{2} \sqrt{B n e} h_{n-1}
   h_n}{E_i+m}\right)\\ \nonumber
.\eeqa   
\end{widetext}

This is a very rich equation. It shows all the interactions between our trapped electron and another field whose potential is $A_\si$. For example, time dependent $A_0$ gives rise to an electric field, so the above shows a spin flip may occur due to an electric field (it also shows this term is $v/c$ times the terms involving the magnetic field). All this is expected, but this formalism displays the actual terms explicitly, and can also be used to derive selection type rules, which is why the Dirac equation is necessary.

 For the case under study we shall consider a circularly polarized wave so that we are investigating a spin-spin interaction, spin of the electron and spin of the electromagnetic wave. The potential for such a wave is

\beq
A_\mu=\frac{ cF}{\om}\{0, \sin(kz-\om t), r \cos(kz-\om t),0\}
\eeq
where $0  \leq r \leq 1$ represents the degree of polarization with $r=0$ being a plane polarized wave and $r=1$ a circularly polarized wave.

Now we set
\beq
S_{fi}=S_{fi}^{(1)}+S_{fi}^{(2)}
\eeq
where

\beqa\label{(1)}
S_{fi}^{(1)}=
-\frac{iC_fC_i}{2L}\int d^4x 
e^{i(E_ft-p_x^fx-p_z^fz)}e^{-\xi^2}\\ \nonumber
M^{(1)}
(e^{i(kz-\om t)}-e^{-i(kz-\om t)})
e^{-i(E_it-p_x^ix-p_z^iz)}
\eeqa
and

\beqa\label{(2)}
S_{fi}^{(2)}=
-\frac{irC_fC_i}{2L}\int d^4x 
e^{i(E_ft-p_x^fx-p_z^fz)}e^{-\xi^2}\\ \nonumber
M^{(2)}
(e^{i(kz-\om t)}+e^{-i(kz-\om t)})
e^{-i(E_it-p_x^ix-p_z^iz)}
\eeqa
where
\beq
M^{1,2}=\overline u_f \gamma^{1,2}u_i
\eeq
and where $\overline u_f =u_f\dag\gamma_0$
We further write

\beq
S_{fi}^{(1)}=S_{fi}^{(1+)}+S_{fi}^{(1-)}
\eeq
and     

\beq
S_{fi}^{(2)}=S_{fi}^{(2+)}+S_{fi}^{(2-)}
\eeq
so that
\beq
S_{fi}=S_{fi}^{1+}+S_{fi}^{1-}+S_{fi}^{2+}+S_{fi}^{2-}
.\eeq

Consider the $x$ and $z$ integrals. These will yield
terms defined as follows

\beq
L_x\equiv\int dx e^{-i(p^f_x-p^i_x)}
\eeq
and

\beq
L_z^\pm\equiv\int dz e^{-i(p^f_z-p^i_z \pm k)}
\eeq
so that

\beqa
S_{fi}=-\frac{C_fC_i}{2L}L_xL_z^-\int dydt e^{i\de^+t}e^{-\xi^2}M^1\\ \nonumber
+\frac{C_fC_i}{2L}L_xL_z^+\int dydt e^{i\de^-t}e^{-\xi^2}M^1\\ \nonumber
+\frac{C_fC_i}{2L}L_xL_z^-r\int dydt e^{i\de^+t}e^{-\xi^2}M^2\\ \nonumber
+\frac{C_fC_i}{2L}L_xL_z^+r\int dydt e^{i\de^-t}e^{-\xi^2}M^2.\\
\eeqa

\no To proceed, we define

\beq
L_t^\pm=c\int dt e^{i\de^\pm t}
\eeq
where we integrate from minus infinity to the present time $t$,
and define the probability of transition:

\beq\label{w}
W=\int |S_{fi}|^2 d\rho
\eeq
where the density of states in $d\rho=(L_x/2\pi\hbar)(L_z/2\pi\hbar)dp_xdp_z$.

The result is
\beq\label{rrf}
W=\left(\frac{p_x^i}{E_i+m}-\frac{p_x^i+\hbar k}{E_f+m}\right)
\kappa \left(\frac{\sin\Delta}{\Delta}\right)^2(1+r^2)
.\eeq
Now let $1/(E_i+mc^2)\approx 1/(E_f+mc^2)=1/2mc^2$.
Let us define $E_f-E_i=\hbar \om_0$ and $\Delta=(\om_0-\om)t/2$, then (\ref{w}) yields
\beq
W=\kappa \left(\frac{\sin\Delta}{\Delta}\right)^2(1+r^2)
\eeq
where $\kappa = \mu_b E/\hbar$ and $\mu_b=e\hbar/2mc$.

With $r=0$ we recognize this as the Rabi formula.\cite{rabi} The effects of circular polarization are evident, for $r=1$ the probability of spin flip doubles. Physically this makes sense if we think about a circularly polarized wave as the sum of two plane waves.

The result (\ref{rrf}) with no approximations may be viewed as the relativistic Rabi formula.

In summary, for an electron trapped in a uniform field, we have considered the interaction of it and an electromagnetic wave. Thus was done using the covariant formalism of the scattering matrix and the solutions to the Dirac equation for an electron in a uniform magnetic field.

In this formalism we do not have to put the interaction in by hand, all effects come out of scattering matrix as shown in (\ref{rich}). The case of an elliptically polarized wave is then considered which results in a relativistic Rabi formula, and the lowest energy result is displayed.

\ed
S_{fi}=\frac{C_iC_f(2\pi)^2d\sqrt{\pi}\de(p_x^f-p_x^i)\de(p_z^f-p_z^i)}{2L\sqrt{eB}}I
\eeq